# Transverse Recoil Imprinted on Free-Electron Radiation

Xihang Shi[1], Lee Wei Wesley Wong[2], Sunchao Huang[2], Liang Jie Wong[2] and Ido Kaminer[1†]

[1]*Solid State Institute and Faculty of Electrical and Computer Engineering, Technion – Israel Institute of Technology, Haifa 3200003, Israel*

[2]*School of Electrical and Electronic Engineering, Nanyang Technological University, 50 Nanyang Avenue, Singapore 639798, Singapore*

[†]Corresponding Author: *kaminer@technion.ac.il*

## Abstract

Phenomena of free-electron X-ray radiation are treated almost exclusively with classical electrodynamics, despite the intrinsic interaction being that of quantum electrodynamics. The lack of quantumness arises from the vast disparity between the electron energy and the much smaller photon energy, resulting in a small cross-section that makes quantum effects negligible. Here we identify a fundamentally distinct phenomenon of electron radiation that bypasses this energy disparity, and thus displays extremely strong quantum features. This phenomenon arises when free-electron transverse scattering occurs during the radiation process, creating entanglement between each transversely recoiled electron and the photons it emitted. This phenomenon profoundly modifies the characteristics of free-electron radiation mediated by crystals, compared to conventional classical analysis and even previous quantum analysis. We also analyze conditions to detect this phenomenon using low-emittance electron beams and high-resolution X-ray spectrometers. These quantum radiation features could guide the development of compact coherent X-ray sources facilitated by nanophotonics and quantum optics.

## Introduction

Despite the fundamental electron-photon interaction being at the core of quantum electrodynamics, the broad range of free-electron radiation effects is to a large part captured using just classical electrodynamics[1–3]. The underlying reason for the lack of quantumness is the intrinsic weakness of electron-photon interactions: small cross-section and negligible inelastic electron recoil by photon emission. Quantum effects appear in the dynamics of electrons once the inelastic recoil energy becomes comparable to or larger than the uncertainty in the electron energy[4–6]. However, despite the electron dynamics becoming quantum, the radiation itself remain classical, and its dispersion is not affected. The general conception is that for quantum effects to also appear in electron radiation, the inelastic electron recoil must become comparable to the actual kinetic energy of the electron. The first observation of such a quantum correction[7] was believed to finally complete the picture of quantum effects induced by electron recoil.

Breaking these expectations, we identify a fundamentally distinct phenomenon of electron radiation – showing features of elastic recoil. While electron radiation is necessarily an inelastic process, we find that additional elastic scattering features can occur simultaneously and create new pathways for electron radiation that are intrinsically quantum. Elastic scattering effects are common in electron diffraction[8–13], involving transverse changes of the electron momentum that preserve the electron coherence. Such effects were not previously related to electron radiation. For context, the inelastic process of electron radiation typically involves recoil in the longitudinal electron momentum, which is the dominant momentum component. In contrast, the elastic recoil alters the transverse electron momentum, which is typically substantially smaller than the longitudinal one and thus much more sensitive to small changes. Consequently, we use the term transverse recoil correction to distinguish the quantum correction proposed and analyzed below (Fig. 1).

The transverse recoil is inherently tied to the wave nature of electrons, which has been known for almost one hundred years, clarified and promoted by the work of Louis de Broglie[14]. However, the connection of electron transverse recoil to electron radiation emission has generally been overlooked, probably due to a lack of awareness in those early days of the role of entanglement in electron radiation. In recent years, pioneering works in free-electron quantum optics have revealed the intrinsic entanglement underlying the processes of free-electron radiation[15–20], spawning the field of free-electron quantum optics[21–27]. Below, we show that such entanglement is involved in the transverse (elastic) recoil effects that alter the inelastic electron radiation processes.

We note that the term "free electrons" is widely adopted to characterize a beam of electrons after an initial acceleration stage. Despite the designation "free", these electrons often undergo interactions with external electromagnetic fields or with various media.

In the X-ray regime of free-electron radiation, recent works predicted quantum corrections arising from entanglement created during the inelastic electron scattering responsible for the photon emission[7]. Another type of quantum correction was predicted to shift the dispersion relation of the radiation[28–30] and recently observed in the X-ray regime[7]. Nevertheless, the effects of transverse recoil on radiation have remained elusive.

Here we show that the transverse recoil could be imprinted on the radiation spectra. In this novel phenomenon of free-electron-driven X-ray radiation, a crystalline material mediates a coherent interaction[31], in which the quantum nature of both the electron and the emitted photon play a crucial role. This quantum radiation phenomenon is surprising as it occurs for parameters expected to only yield classical physics, in regimes analyzed classically for a century. We present concrete examples that can be implemented in current experimental setups for X-ray radiation from electrons[7,32,33], as in electron microscopes equipped with X-ray spectrometers.

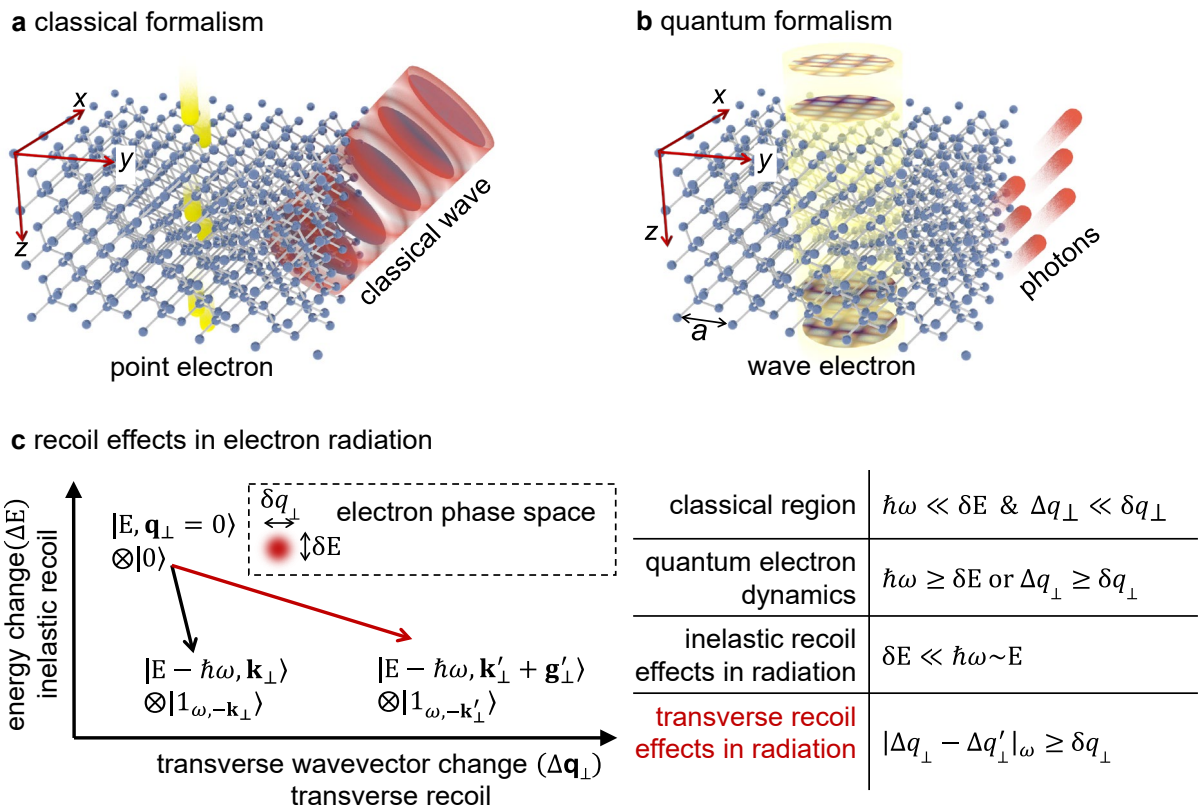

**Fig. 1 Comparing free-electron-driven coherent X-ray radiation (CXR) in the classical and quantum formalisms.** CXR is excited by the coherent interaction of electrons traversing a crystal, such as a silicon film. (**a**) In the classical formalism, the electron is modeled as a point charge, and the radiation as classical waves. (**b**) In the quantum formalism, the electron is represented by a wave, while the radiation by discrete photons. (**c**) Mapping the recoil effects in electron radiation. The axes represent the inelastic and transverse recoil on the electron, defined by the change in its energy and transverse momentum, respectively. The electron-photon joint states are defined by the energy and transverse wavevector of both electrons and photons. In the classical domain, the inelastic recoil ($\Delta E = \hbar\omega$) and the transverse recoil ($\Delta q_\perp$) experienced by the electron are minor relative to the phase space uncertainties of the electron. Quantum corrections to electron dynamics become significant when the recoils approach the electron uncertainties. Yet, inelastic recoil effects in radiation require an even larger inelastic recoil, comparable to the electron kinetic energy[7]. The strongest influence of transverse recoil on the radiation appears for photon emission that is accompanied by different transverse wavevector changes ($\Delta q_\perp$ and $\Delta q'_\perp$), with the difference exceeding the electron uncertainty $\delta q_\perp$.

We show how the signature of transverse recoil on electron radiation can arise in several mechanisms of X-ray emission from free electrons, which have been largely seen as distinct radiation phenomena and can now be studied using a uniform framework[34–37]. To explain this universal behavior, we develop a framework unifying the mechanisms of parametric X-ray radiation (PXR)[38–40], coherent Bremsstrahlung (CB)[31,41,42], and higher-order electron-driven X-rays[43]. We denote the unified phenomenon as coherent X-ray radiation (CXR). To account for both electron inelastic and transverse recoils, we adopt a quantum formalism comprising both the particle nature of the photons and the wave nature of the electrons (Fig. 1b). We show in what aspects CXR is affected by the inelastic and transverse recoils, and where it breaks the predictions involving the nonrecoil approximation and inelastic recoil corrections (Fig. 1a).

In the process of free-electron radiation, the electron experiences inelastic recoil, losing energy to the emitted photons. Meanwhile, the electron undergoes transverse recoil due to different transverse reciprocal lattice vector components, as illustrated in Fig. 1c. The quantum corrections due to inelastic recoil effects in radiation[7,44] are known to shift the spectra relative to classical predictions. In contrast, we find that the transverse recoil effect does not simply shift the radiation spectra, but instead splits the spectra, with each split corresponding to different transverse recoil pathways. Direct estimation of the transverse-recoil-induced shifts and splits in the radiation spectra can be derived in the context of momentum-energy conservation, as detailed in the Methods. However, a full quantum theory following the Fermi's golden rule is required to analyze the radiation intensity. We note that this splitting effect not only introduces additional shifts but also significantly diminishes the peak intensities of the CXR.

Interestingly, the polarization of CXR is strongly affected by the transverse recoil. PXR and CB are known to be linearly polarized when excited by ultra-relativistic electrons[35,45–48]. However, when excited by semi-relativistic electrons, the conventional classical formalism predicts mostly (mixed) unpolarized light because multiple radiation modes that arise from different recoil pathways cannot be separated. We show that the recoil-induced splitting separates the different radiation modes associated with different electron recoil pathways. Consequently, CXR can become strongly (linearly) polarized when the radiation is entangled with electrons with a singular transverse recoil pathway.

Our findings pave the way for shaping and enhancing free-electron radiation via tailored incident electron wavefunctions, contributing to the development of compact, coherent X-ray sources with desired quantum-optical characteristics.

## Results

### The big picture of quantum effects in free-electron radiation

In the processes of coherent free-electron radiation, electrons experience recoil attributed to longitudinal and transverse momentum variation[12,49–52], as shown in Fig. 1c. These two recoil pathways are disregarded[1–3] in the classical descriptions of electron radiation[1–3,53–63] based on Maxwell equations. Under the nonrecoil approximation, the radiation power spectrum is independent of the single electron wavefunction[64,65].

Electron recoil can be significantly enhanced through interaction with external laser fields. Earlier experiments[11,66] have demonstrated such interaction in the form of a Kapitza-Dirac effect[9] mediated by the ponderomotive force of free-space light. Studies in the last decade rely on electrons interacting with evanescent light fields, developing into a new field in electron microscopy: photon-induced near-field electron microscopy (PINEM)[4–6]. In such cases, the inelastic electron recoil is larger than the electron energy uncertainty, and it thus alters the electron dynamics. The quantum wave nature of the electron dynamics led to the observation of a plethora of quantum effects such as multiple quanta of photon absorption/emission[51,52], free-electron quantum walk and Rabi oscillations[67], and imprinting photon quantum statistics on the electron energy spectrum[26].

In sharp contrast to these observations of recoil effects on electron dynamics, it is generally believed that only when the inelastic energy loss is comparable to the electron kinetic energy, will detectable quantum effects occur in the radiation itself[1,28,29,65]. This much stricter condition was much harder to accomplish, yet it was recently reached in experiments[7] in the X-ray regime.

The role of electron inelastic recoil is also pivotal in free-electron lasers (FELs), influencing electron microbunching, gain[68,69], and photon statistics[70]. The electron's inelastic recoil (relative to its energy uncertainty) also determines whether an FEL operates in the classical or quantum region[71,72].

Nonetheless, the essential features of currently operational FELs, such as their gain, could all be explained using the classical framework (the quantum recoil is seen as a correction[69,73]). In contrast, the transverse recoil effect that we propose and analyze in this work is characterized by a shift in transverse momentum. This effect does not have an analogue in current FEL research because the transverse recoil does not significantly affect existing FEL operation.

The transverse recoil effect that we identify in this work differs from all the above by its ubiquitous nature. This effect is indirectly implied in previous studies of CB and channeling radiation from dressed electrons[44,74–76]. There, the elastic scattering is not considered as a recoil correction but as a complete redressing of the electron state. Modulated by the periodic atomic potential, the electrons traversing the crystals follow a Bloch-type dispersion[44,77], which exhibits band gaps. In contrast, this work shows that the effects of transverse electron recoil (or electron dressing) on radiation are not limited to specific cases of dressed electrons in CB, but have a much stronger effect on PXR. We capture the combined effect of PXR and CB using a combination of first and second-order perturbation theory. Most importantly, the transverse recoil effects on PXR are dominant because they occur with free electrons following first-order perturbation theory. This quantum effect by free electrons is of great significance as it applies directly to a wide range of additional electron radiation phenomena, such as Smith-Purcell radiation[54,55,57,60], Cherenkov radiation[28,53,59,61], transition radiation[59,78], etc[58]. As we show below, the quantum effect that we predict differs from previous studies so far, and it is also more accessible in current experimental setups.

**The classical framework of coherent X-ray radiation**

In the classical framework (Fig. 1a), the radiation is generated by a point charge subject to the Newton-Lorentz equation and the Maxwell equations[37,79,80]. The photon emission rate, per unit angular frequency $\omega$ and per unit solid angle $\Omega$, from a point charge moving in a crystal with an arbitrary dispersion, is given by the following compact form:

$$\frac{d^2N}{d\omega d\Omega} = \frac{\alpha\omega}{(2\pi c)^2} \sum_\sigma \left| \int_{-\infty}^{\infty} dt\, \mathbf{v}_e(t) \cdot \mathbf{E}_{\mathbf{k},\sigma}(\mathbf{r}_e(t), \omega) e^{-i\omega t} \right|^2, \qquad (1)$$

where $\alpha$ is the fine-structure constant, $c$ is the vacuum light speed, $\mathbf{k}$ and $\sigma$ are the wavevector and polarization of the emitted photon. $\mathbf{v}_e(t)$ is the time-dependent velocity of the free electron, and $\mathbf{E}_{\mathbf{k},\sigma}(\mathbf{r}_e(t), \omega)$ is the quantized electrical field inside the crystal, which implicitly depends on time through the electron trajectory $\mathbf{r}_e(t)$. The derivation of $\mathbf{E}_{\mathbf{k},\sigma}(\mathbf{r}_e(t), \omega)$ is detailed in the Methods.

The trajectory $\mathbf{r}_e(t) = \mathbf{r}_0 + (v_0\hat{z} + \delta\mathbf{v}(t))t$ is determined by solving the Newton-Lorentz equation within a periodic potential $\phi(\mathbf{r})$. For energetic electrons used to produce X-ray radiation, it is possible to solve for $\delta\mathbf{v}(t)$ as a first-order perturbation relative to the paraxial dynamics $v_0\hat{z}$. By neglecting $\delta\mathbf{v}(t)$ in Eq. (1), one could obtain the emission rate of PXR, which accounts for the majority of the CXR in most cases when the electron is 300 keV and below. The initial position of the electron $\mathbf{r}_0$ is typically averaged over the crystalline unit cell in the simulation.

The framework above already contains certain quantum properties affecting the field mode in the crystal, and already considers the electron modulation by the periodic potential. However, this theoretical framework does not account for the electron wave nature and its dynamical evolution in the crystal, which we show below to be of substantial influence on all the radiation properties.

**The quantum framework of coherent X-ray radiation**

The quantum description of the free-electron-driven CXR process in crystals is based on the Hamiltonian $H = H_\text{e} + H_\text{ph} + H_\text{e,ph} + H_\text{e,c}$, where $H_\text{e}$ and $H_\text{ph}$ are the unperturbed Hamiltonians of free electrons and free photons, whereas $H_\text{e,ph}$ and $H_\text{e,c}$ are the interaction Hamiltonians of free electrons with the quantized vector potential[81,82] and the Coulomb potential[31,81] inside crystals, respectively. Their full representations are

$$H_\text{e} = \int d\mathbf{r}\ \psi^\dagger(\mathbf{r})(c\boldsymbol{\alpha}\cdot\mathbf{p} + \boldsymbol{\beta} mc^2)\psi(\mathbf{r}),$$

$$H_\text{ph} = \sum_{\mathbf{k},\sigma} \hbar\omega a_{\mathbf{k},\sigma}^\dagger a_{\mathbf{k},\sigma},$$

$$H_\text{e,ph} = -ce \int d\mathbf{r}\ \psi^\dagger(\mathbf{r})\boldsymbol{\alpha}\cdot\widehat{\mathbf{A}}(\mathbf{r})\psi(\mathbf{r}),$$

$$H_\text{e,c} = e \int d\mathbf{r}\ \psi^\dagger(\mathbf{r})\phi(\mathbf{r})\psi(\mathbf{r}),$$

(2)

where $\psi(\mathbf{r}) = \frac{1}{\sqrt{V}}\sum_{\mathbf{q},s} b_{\mathbf{q},s}\mathbf{u}_{\mathbf{q},s}e^{i\mathbf{q}\cdot\mathbf{r}}$ is the position space annihilation operator of free electrons, $\mathbf{q}$ and $s$ are the wavevector and spin of each electron state, $b_{\mathbf{q},s}$ is the annihilation operator of the electron wavevector state, $\mathbf{u}_{\mathbf{q},s}$ is the normalized four-component spinor satisfying $\mathbf{u}_{\mathbf{q},s}^\dagger \mathbf{u}_{\mathbf{q},s} = 1$, $a_{\mathbf{k},\sigma}^\dagger$ and $a_{\mathbf{k},\sigma}$ are the photonic creation and annihilation operators, $e$ is the electron charge, $m_\text{e}$ is the electron mass, $\mathbf{p} = -i\hbar\nabla$ is the momentum operator, and $\boldsymbol{\alpha} = \begin{bmatrix} 0 & \boldsymbol{\sigma} \\ \boldsymbol{\sigma} & 0 \end{bmatrix}, \boldsymbol{\beta} = \begin{bmatrix} \mathbf{I} & 0 \\ 0 & -\mathbf{I} \end{bmatrix}$ are the Dirac matrices defined in terms of the two-dimensional Pauli matrix $\boldsymbol{\sigma}$ and the identity matrix $\mathbf{I}$. The second quantized vector potential is denoted by $\widehat{\mathbf{A}}(\mathbf{r})$ and the scalar (screened) Coulomb potential inside the crystal is denoted by $\phi(\mathbf{r})$. The complete representations of them are given by Eqs. (4) and (5) in the Methods.

We adopt a joint electron-photon state to account for the entanglement of each electron to its emitted photons[15,16,20] arising due to the electron recoil (both transverse and longitudinal). Without loss of generality, we start by considering an initial electron superposition state $|\text{ini}\rangle = \sum_{\mathbf{q},s} \frac{1}{\sqrt{V}} \varphi_s(\mathbf{q})|1_{\mathbf{q}s}\rangle$ composed of multiple electron wavevectors $|1_{\mathbf{q}s}\rangle$, where $\varphi_s(\mathbf{q})$ is the amplitude normalized by $\sum_{\mathbf{q},s} \frac{1}{V}|\varphi_s(\mathbf{q})|^2 = 1$. The differential emission rate of an output photon of wavevector $\mathbf{k}$ and polarization $\sigma$ is

$$\begin{aligned}
\frac{d^2 N}{d\omega d\Omega} &= \frac{\omega^2}{c^3(2\pi)^3} \sum_{\mathbf{q}_f s_f \sigma} \left| \sum_{\mathbf{q}_i s_i \mathbf{g}} \varphi_{s_i}(\mathbf{q}_i) \left( M'^{\text{PXR}}_{\mathbf{q}_i s_i \mathbf{g} \to \mathbf{q}_f s_f, \mathbf{k}\sigma} + M'^{\text{CB}}_{\mathbf{q}_i s_i \mathbf{g} \to \mathbf{q}_f s_f, \mathbf{k}\sigma} \right) \delta_{\mathbf{q}_i+\mathbf{g},\mathbf{q}_f+\mathbf{k}} \right|^2 \\
&= \frac{\omega^2}{c^3(2\pi)^3} \sum_{s_f \sigma} \sum_{\substack{\mathbf{q}_i s_i \mathbf{g} \\ \mathbf{q}_j s_j \mathbf{h}}} \varphi_{s_i}(\mathbf{q}_i) \varphi^*_{s_j}(\mathbf{q}_j) M'^{\text{CXR}}_{\mathbf{q}_i s_i \mathbf{g} \to \mathbf{q}_f s_f, \mathbf{k}\sigma} \left( M'^{\text{CXR}}_{\mathbf{q}_j s_j \mathbf{h} \to \mathbf{q}_f s_f, \mathbf{k}\sigma} \right)^* \delta_{\mathbf{q}_i+\mathbf{g},\mathbf{q}_j+\mathbf{h}},
\end{aligned} \quad (3)$$

where $\omega = |\mathbf{k}|c$ and $\sigma$ are the angular frequency and polarization of photons, $\mathbf{q}_i$, $s_i$, $\mathbf{q}_j$, $s_j$ ($\mathbf{q}_f$, $s_f$) run over initial (final) electron wavevector states, $\mathbf{g}$ and $\mathbf{h}$ are the reciprocal lattice vectors. $M'^{\text{PXR}}_{\mathbf{q}_i s_i \mathbf{g} \to \mathbf{q}_f s_f, \mathbf{k}\sigma}$ and $M'^{\text{CB}}_{\mathbf{q}_i s_i \mathbf{g} \to \mathbf{q}_f s_f, \mathbf{k}\sigma}$ are the respective scattering elements of PXR and CB processes from the initial electron state $|1_{\mathbf{q}_i s_i}\rangle$ to the final electron-photon joint state $|1_{\mathbf{q}_f s_f}\rangle \otimes |1_{\mathbf{k}\sigma}\rangle$, with the momentum conservation represented by the Kronecker delta implicitly enforced. $|1_{\mathbf{k}\sigma}\rangle$ is the Fock state of radiation with one photon, and $M'^{\text{CXR}}_{\mathbf{q}_i s_i \mathbf{g} \to \mathbf{q}_f s_f, \mathbf{k}\sigma} = M'^{\text{PXR}}_{\mathbf{q}_i s_i \mathbf{g} \to \mathbf{q}_f s_f, \mathbf{k}\sigma} + M'^{\text{CB}}_{\mathbf{q}_i s_i \mathbf{g} \to \mathbf{q}_f s_f, \mathbf{k}\sigma}$. Details of the derivation leading to Eq. (3) and the representation of the scattering elements are given in the Methods and Supplementary Section 3.

**Transverse recoil effects on the spectral and polarization characteristics of the radiation**

In this section, we demonstrate that CXR depicted by the quantum formalism sharply contrasts with the classical descriptions. We find that the angular spectral distribution of CXR features numerous spectral peaks, which are split, shifted, and reduced in intensity, as compared to the classical predictions. Additionally, photons at different peaks exhibit complex polarization characteristics, in contrast to the unpolarized nature predicted by the classical framework.

In this study, we utilize a plane wave electron, which is a common approximation for a transversely delocalized electron (e.g., a wide Gaussian beam) consisting of momentum components that are tightly focused around a central value (see Supplementary Section 6). In Fig. 2, we present the radiation spectrum from electrons with kinetic energies of 30 keV, 300 keV, and 5 MeV traversing 30 nm silicon films along the normal. In what follows, we introduce two terms, quantum CXR and classical CXR, to distinguish between describing CXR in the quantum formalism, as described by Eqs. (2) and (3), and in the classical formalism, by Eq. (1). By comparison, we note that the angular spectral distribution of quantum CXR is more complex than those of classical CXR. The classical CXR follows the linearized dispersion $\omega = (\mathbf{k} - \mathbf{g}) \cdot v_0 \hat{z}$, which does not account for the full dispersion of the electron. Consequently, the predicted radiation frequencies are determined by the longitudinal periodicity (along the electron trajectory) of the crystal structures, similar to Smith-Purcell radiation[54,55,57,60]. However, the spectra of quantum CXR, in certain radiation direction associated with the same longitudinal periodicity, exhibit multiple peaks. By increasing the electron energy, the multiple peaks in the spectrum approach each other, gradually converging to coincide with the spectrum of classical CXR. This spectral behavior agrees with the analysis from energy-momentum conservation in the Methods.

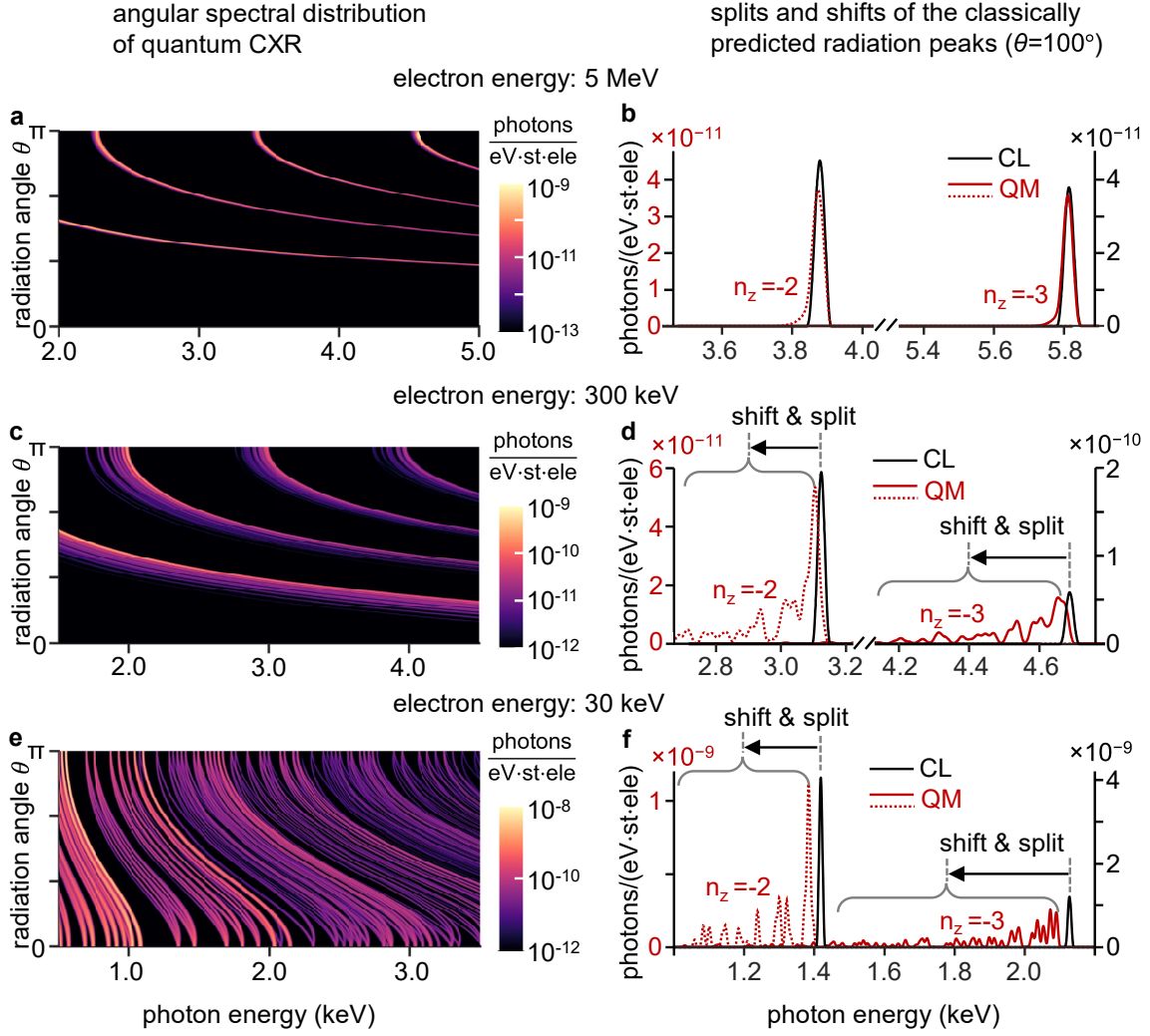

**Fig. 2 Transverse recoil effects in the spectra of CXR.** (**a**) (**c**) and (**e**) plot the angular spectral distribution of quantum CXR. (**b**) (**d**) and (**f**) compare the spectra of quantum CXR and classical CXR at the radiation angle $\theta = 100°$. The curves in (**a**) (**c**) (**e**) and the radiation peaks of quantum CXR in (**b**) (**d**) (**f**) are associated with different reciprocal lattice vectors $\mathbf{g} = (n_x, n_y, n_z)\frac{2\pi}{a}$, where $a$ is the lattice constant of silicon. In contrast, the frequency of the classical CXR is solely determined by the longitudinal reciprocal lattice vectors labeled by $n_z$. Therefore, the quantum CXR peaks split and shift toward lower photon energies, as shown in (**d**) and (**f**). By increasing the electron energy, the multiple peaks gradually converge to coincide with the classically predicted peaks, as shown in (**a**) and (**b**). The electrons move along the $z$ direction with kinetic energies of 5 MeV, 300 keV, and 30 keV.

During the radiation process, the electrons are scattered by the reciprocal lattice of the crystal structure, as represented by the Kronecker delta in Eq. (3). The transverse ($\mathbf{g}_\perp$) and longitudinal ($\mathbf{g}_\parallel$) reciprocal lattice vector components are associated with the transverse and longitudinal electron recoils, respectively. This is due to the fact that $\mathbf{g}_\perp$ and $\mathbf{g}_\parallel$ bring distinct modifications upon the electron states, causing the electron to deflect and decelerate, respectively. Realizing that the emitted photons and final electrons are entangled, the radiation spectra are associated with various $\mathbf{g}_\perp$ and $\mathbf{g}_\parallel$. Analysis of energy-momentum conservation shows that $\mathbf{g}_\perp$ and $\mathbf{g}_\parallel$ contribute quantum corrections on the same order to the emitted photon frequency (see Methods). Thus, we conclude that the quantum characteristics of CXR arise from a combination of longitudinal and transverse recoil effects.

Despite the inelastic nature of the spontaneous radiation process, inelastic recoil can be negligible in its effect on the radiation, and in fact, this is usually the case for most experimental scenarios. In such cases, transverse recoil becomes the dominant factor for quantum effects on radiation. This situation is demonstrated in Figs. 2c-2f and by additional examples in Supplementary Section 4. Figs. 2(b), (d) and (f) present the emission spectra at a fixed radiation direction. These plots correspond to an X-ray spectrometer that captures a wide spectral range while located at a fixed angle. To highlight the transverse recoil effects, the next section presents the angular distribution of photons at specific fixed energies.

We note that in CXR from van der Waals (vdW) materials, such as $WSe_2$ [33] and graphite[7], the radiation peaks corresponding to $\mathbf{g}_\perp = 0$ are dominant, while those with $\mathbf{g}_\perp \neq 0$ are significantly weaker (refer to Supplementary Section 5 for details). This explains why the transverse-recoil-induced spectral splits were not observable in previous experiments[7,33]. The quantum effects will become evident for CXR from silicon once locating the spectrometer at a small angle relative to the electron beam direction. In larger radiation directions, as shown in Figs. 2(b), (d) and (f), distinguishing the split peaks demands higher angular and energy resolution. However, measuring at small radiation angles simplifies the requirements from the spectrometer, making detection feasible using existing experimental platforms, as evidenced in Fig. 2e. For a detailed analysis, refer to Supplementary Section 10.

The unique polarization pattern of CXR and its strong deviation from the classical prediction is also attributed to the transverse recoil. Interestingly, the classically predicted PXR and CB from ultra-relativistic electrons are linearly polarized when scattered by a single reciprocal lattice vector[45–48]. However, once considering scattering by multiple transverse reciprocal lattice vectors[37,80], the radiation from semi-relativistic electrons becomes almost unpolarized (black dashed curve Fig. 3b). We find that this unpolarized nature arises from disregarding the transverse recoil in the classical theories. In the quantum formalism, radiation peaks could be associated with different transverse recoil pathways (Fig. 3a). Once the radiation at a certain angle and energy is connected to an individual recoil pathway, its polarization is strongly linear (curve II in Fig. 3b). Whenever the radiation at a certain angle and energy is connected to two transverse recoil pathways, photons of different polarizations combine incoherently, resulting in a mixed state of unpolarized light.

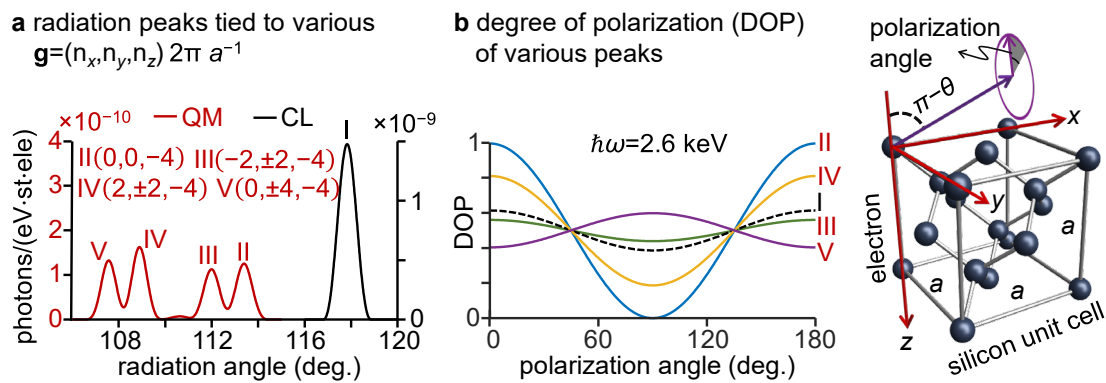

**Fig. 3 Transverse recoil effects in the polarization of monochromatic CXR.** (**a**) Quantum CXR peaks (II-V) are associated with different transverse recoil pathways, i.e., transverse reciprocal lattice vectors ($\mathbf{g}_\perp$). The classical CXR peak (I) is associated with all possible transverse recoil pathways. (**b**) The degree of polarization (DOP) of radiation at photon energy $\hbar\omega = 2.6$ keV. The classical approach predicts nearly unpolarized light, as represented by the black dashed curve. The quantum CXR peak (II), which is associated with an individual elastic recoil, is linearly polarized. The spectra and the DOP are analyzed with radiation in the $x - z$ plane, where the polarization angle is set to zero. The inset on the right illustrates the silicon unit cell, the radiation angle $\theta$ in the $x - z$ plane, and the polarization angle. The electron energy is 30 keV.

**Transverse recoil effects in the angular distribution favoring the experimental detection**

We compare the angular distributions of the monochromatic X-rays in Fig. 4. The azimuthal angle $\varphi$ and the polar angle $\theta$ represent the radiation angle with respect to the $x - z$ plane and the electron velocity, respectively. Under normal incidence, the azimuthal distribution of quantum CXR, as shown in Fig. 4c, clearly exhibits the same rotational symmetry as the transverse reciprocal lattices. In contrast, the azimuthal distribution of classical CXR is approximately rotationally symmetric, as depicted in Fig. 4a.

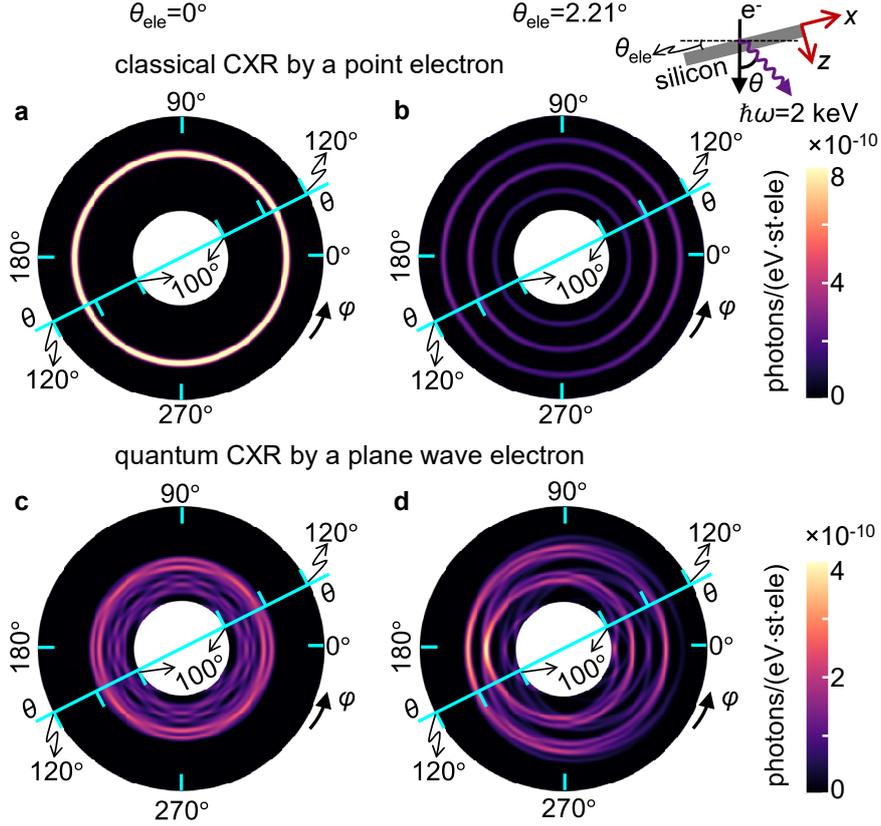

**Fig. 4 Transverse recoil effects in the angular distribution of monochromatic CXR.** The azimuthal angle $\varphi$ and the polar angle $\theta$ represent the radiation angle with respect to the $x-z$ plane and the electron velocity, respectively. The tilt angle $\theta_{\text{ele}}$ extends between the $z$-axis of the crystal (as sketched in Fig. 3) and electron velocity. When $\theta_{\text{ele}} = 0$, the azimuthal distribution of classical CXR (**a**) is approximately rotationally symmetric. However, the azimuthal distribution of quantum CXR (**c**) exhibits the same rotational symmetry as the transverse reciprocal lattices, which are composed of square unit cells. When the sample is tilted, the angular distribution pattern of classical CXR (**b**) splits and becomes azimuthally anisotropic, decreasing the peak intensities. However, the quantum CXR (**d**) becomes enhanced at specific peaks along $\varphi = 0°$ and $180°$. The photon energy is 2.0 keV. The tilt angle in (**b**) and (**d**) is $\theta_{\text{ele}} = \tan^{-1}\left(\frac{q_x}{q_z}\right) = 2.21°$, where $q_x = \frac{6\pi}{a}$ and $q_z$ are the $x$ and $z$ components of the wavevector of a 30 keV electron. The inset on the top-right illustrates the sample tilt angle $\theta_{\text{ele}}$ and the radiation angle $\theta$.

The peaks of classical CXR also split by tilting the crystal relative to the electron velocity (Fig. 4b). The radiation direction of classical CXR is determined by the dispersion $\omega = (\mathbf{k} - \mathbf{g}) \cdot \mathbf{v}$. When the electron is normally incident along the $z$-axis, the $z$ component reciprocal lattice vector ($g_\parallel = g_z$) determines the radiation dispersion. We then orient the $xyz$ frame based on the crystal structure as depicted in the inset of Fig. 3. When the sample is tilted by an angle $\theta_{\text{ele}}$, the longitudinal reciprocal lattice vector $g_\parallel = g_z \cos\theta_{\text{ele}} - g_x \sin\theta_{\text{ele}}$. Therefore, different in-plane components $g_x$ of the reciprocal lattice vectors result in the split radiation pattern in Fig. 4b. The peak intensities in the classical theory, correspondingly, always decrease by tilting the sample.

On the contrary, the intensities of specific peaks of quantum CXR increase when the in-plane periodicity of a tilted sample is phase-matched with the electron wavevector. For instance, the phase of the plane wave electron becomes $\exp\left(-i\frac{6\pi}{a}x + iq_z z\right)$ in Fig. 4, with the tilting angle $\theta_{\text{ele}} = \tan^{-1}\left(\frac{6\pi/a}{q_z}\right)$, where $q_z$ is the $z$ component of the electron wavevector. The quantum CXR in Fig. 4d exhibits brighter monochromatic peaks in the plane $\varphi = 0°$ and $180°$. The disparities over the angular distribution could facilitate the experimental verification of quantum CXR even without requiring high-resolution angle-resolved detectors.

**Coherently shaped CXR by tailoring electron wavefunctions**

Transverse recoil effects in CXR provide a new route to coherently control free-electron radiation. When the initial electron state in momentum space is tightly focused around a central value, the quantum processes associated with different transverse recoil pathways are *incoherently* summed, i.e., Eq. (2) becomes $\frac{d^2N}{d\omega d\Omega} \propto \sum_{\mathbf{q}_i s_i \mathbf{g}} |\varphi_{s_i}(\mathbf{q}_i) M'^{\text{CXR}}_{\mathbf{q}_i s_i \mathbf{g} \to \mathbf{q}_f s_f, \mathbf{k}\sigma}|^2$, with additional summation over photon polarization and electron spin. However, we can coherently sum different transverse recoil pathways when the electron wavefunction is engineered to have transverse wavevectors focused along different transverse reciprocal lattice vectors. The underlying principle enabling this coherent summation was first applied to bremsstrahlung radiation from an electron superposition state scattered by a single atom[15] and by monolayer 2D materials[83].

Here, we rely on this coherent summation for a 3D bulk crystal, thus capturing the full effect of CB. More importantly, we show that such coherent summation can also be applied for the mechanism of PXR from bulk crystals, thus covering the unified effect of CXR. In such a context, PXR constitutes the primary component of the radiation, with transverse recoil pathways being discretely distributed. For specific electron wavefunctions, various photon emission pathways that correspond to different transverse recoils converge to the same joint electron-photon state. Such configurations allow the amplitudes of the emission pathways to interfere coherently, resulting in the strong enhancement of quantum features arising from transverse recoil.

We demonstrate this principle by a superposition of two plane wave states $|\text{ini}\rangle = \frac{1}{\sqrt{2}}(|1_{\mathbf{q}_1 s}\rangle + e^{i\varphi_{12}}|1_{\mathbf{q}_2 s}\rangle)$, which can be created by a bi-prism or a phase plate in electron microscopy[84,85]. We choose $\mathbf{q}_1$ and $\mathbf{q}_2$ to differ by exactly one transverse reciprocal lattice

vector $\mathbf{g}_\perp$ such that $\mathbf{q}_1 = \mathbf{q}_2 + \mathbf{g}_\perp$. This choice creates interference terms in Eq. (2). Constructive interference can be achieved by tuning the relative phase $\varphi_{12}$. The results are presented in Fig. 5, where the tailored-electron-driven CXR becomes more directional with further separated and brighter peaks along $\varphi = 0°$ and $180°$. Note that the enhancement of the peak intensity results from the coherent interference of the electron superposition state, rather than the electron oblique incidence of each separated state. A quantitative analysis is provided in Supplementary Section 7, comparing the radiation patterns driven by electrons with two plane wave states $|1_{\mathbf{q}_1 s}\rangle$, $|1_{\mathbf{q}_2 s}\rangle$ and their superposition state.

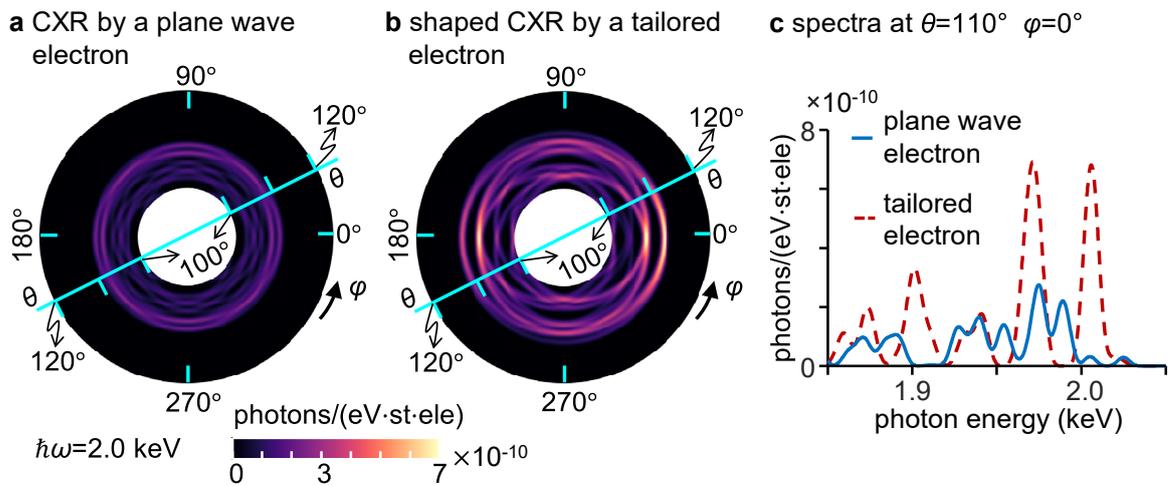

**Fig. 5 Coherently shaped CXR by a tailored electron wavefunction.** Panels (**a**) and (**b**) compare the angular distribution of monochromatic CXR driven by a plane wave electron and a tailored electron, respectively. The tailored-electron-driven CXR has brighter peaks along $\varphi = 0°$ and $\varphi = 180°$, owing to coherent enhancement. Panel (**c**) presents the spectra in one radiation direction, showing that the coherent enhancement applies to a broad frequency range. The superposition state of the tailored electron is $|\text{ini}\rangle = \frac{1}{\sqrt{2}}\left(1_{\left|\mathbf{q}_1=\left(\frac{2\pi}{a},\frac{4\pi}{a},q_z\right)\right\rangle} + 1_{\left|\mathbf{q}_2=\left(\frac{2\pi}{a},-\frac{4\pi}{a},q_z\right)\right\rangle}\right)$. The electron energy is 30 keV.

## Discussion

The electron transverse recoil is intrinsically tied to the wave nature of electrons, as predicted by de Broglie[14] and observed in electron scattering experiments[8]. Six years after de Broglie's work, Kapitza and Dirac proposed that electrons can also be transversely (elastically) scattered by a standing wave of light[9,11]. However, the weak cross-section led to the effect being demonstrated only decades later using a strong laser[10]. More recent works proposed[12] and observed[13] variants on the Kapitza-Dirac effect by the inelastic and elastic interaction with plasmon standing waves. These effects all concern the electron dynamics rather than its radiation. In electron radiation processes, transverse recoil effects have been largely overlooked, possibly due to the common approximation of electron dispersion as linear. It could also be due to a lack of awareness of the implications of having the emitted photons entangled to different pathways of electrons undergoing different transverse recoils.

This work shows that both the transverse and longitudinal recoil of the electrons can result in quantum effects on radiation. We develop a full quantum formalism that captures all X-ray emission processes from electron interactions with crystals, considering the electron quantum wave nature and the second quantization of the radiation. This unified description fully considers the electron recoil and the entanglement between the electron and the radiation. The quantum formalism shows especially strong features for free-electron-driven CXR by semi-relativistic electrons, which is a regime of great promise for devising compact coherent X-ray sources[33,43,63,80,86,87]. We also anticipate that transverse recoil effects on radiation would disappear when the electron spatial coherence is smaller than the crystal transverse periodicity (see Supplementary Section 6). The work presented here is not limited to CXR, but rather extends to free-electron interactions with materials that exhibit lateral nanoscale features[43,88]. In all of these, the quantum transverse recoil effects are expected to arise when the lateral features are smaller than the electron spatial coherence.

The complex, multi-peaked spectra of quantum CXR reflects the nonlinear and spectrally broadband nature of free electrons, which could be used to generate photons with user-defined spectral and spatial properties by leveraging the versatility of the interaction, especially by using designed nanophotonic materials. The peak intensity and separation in energy space could be tuned by tailoring the electron wavefunction. For the case of an incident collimated electron beam, the radiation exhibits narrow frequency-resolved peaks, which could potentially be utilized in X-ray metrology to lock a spectral peak to a desired core-level energy transition. By controlling the relative phase between the peaks, we can also shape the temporal profile of the X-ray pulse, possibly reaching attosecond and even zeptosecond timescales.

In conclusion, our work shows that mostly elastic, transverse recoil plays significant roles in free-electron radiation, despite the inelastic nature of each photon emission process. Disregarding the transverse recoil misses all the effects related to the wave nature of electrons, preventing an accurate analysis of the free-electron interactions with nanostructures. Our findings may spark interest in developing compact, coherent and polarized X-ray sources[43,88], which may facilitate the development of novel methods in X-ray spectroscopy and microscopy[89].

# Methods

## Quantized electrical field, vector potential and screened Coulomb potential

This section outlines the components necessary for calculating CXR from free electrons, which are used both in the classical and in the quantum frameworks.

We define the electric field $\mathbf{E}(\mathbf{r})$ and second quantized vector potential $\hat{\mathbf{A}}(\mathbf{r})$ of a single photon in a bulk crystal[81,90]. Each Fourier component differs from the counterpart in vacuum by a term that is proportional to $\chi_\mathbf{g}$, which is the Fourier component of the susceptibility $\chi(\mathbf{r}) = \sum_\mathbf{g} \chi_\mathbf{g}(\omega) e^{i\mathbf{g}\cdot\mathbf{r}}$, with truncating accuracy up to $O(\chi_\mathbf{g})$. The susceptibility $\chi(\mathbf{r})$ is deduced from the electronic density distribution[90], which can be derived by the screened Coulomb potential as described below. Within the Born approximation, $\mathbf{E}(\mathbf{r})$ and $\hat{\mathbf{A}}(\mathbf{r})$ are

$$\mathbf{E}(\mathbf{r}) = \sum_{\mathbf{k},\sigma} i \sqrt{\frac{\hbar\omega}{2\varepsilon_0 V}} \left( 1 + \sum_\mathbf{g} \chi_{-\mathbf{g}}(\omega) \frac{k^2 \mathbf{I} - (\mathbf{k}-\mathbf{g}) \otimes (\mathbf{k}-\mathbf{g})}{|\mathbf{k}-\mathbf{g}|^2 - k^2} e^{-i\mathbf{g}\cdot\mathbf{r}} \right) \cdot \mathbf{e}_\sigma e^{i\mathbf{k}\cdot\mathbf{r}}$$
$$+ c.c.,$$
$$\hat{\mathbf{A}}(\mathbf{r}) = \sum_{\mathbf{k},\sigma} \sqrt{\frac{\hbar}{2\varepsilon_0 \omega V}} \left( 1 + \sum_\mathbf{g} \chi_{-\mathbf{g}}(\omega) \frac{k^2 \mathbf{I} - (\mathbf{k}-\mathbf{g}) \otimes (\mathbf{k}-\mathbf{g})}{|\mathbf{k}-\mathbf{g}|^2 - k^2} e^{-i\mathbf{g}\cdot\mathbf{r}} \right)$$
$$\cdot \mathbf{e}_\sigma a_{\mathbf{k},\sigma} e^{i\mathbf{k}\cdot\mathbf{r}} + c.c.,$$

(4)

where $\varepsilon_0$ is the vacuum permittivity, $\mathbf{I}$ is a unit dyadic tensor, $\mathbf{e}_\sigma$ is the polarization vector, and $k = |\mathbf{k}|$. The perturbed term, which is proportional $\chi_{-\mathbf{g}}$, characterizes the Bragg scattering of free photons or photonic modes by supplementing one reciprocal lattice vector $\mathbf{g}$. The complete derivation is available in Supplementary Section 2. $\mathbf{E}_{\mathbf{k},\sigma}(\mathbf{r}) = \sum_\mathbf{g} \chi_{-\mathbf{g}}(\omega) \frac{k^2 \mathbf{I} - (\mathbf{k}-\mathbf{g})(\mathbf{k}-\mathbf{g})}{|\mathbf{k}-\mathbf{g}|^2 - k^2} \cdot \mathbf{e}_\sigma e^{i(\mathbf{k}-\mathbf{g})\cdot\mathbf{r}}$ is adopted in Eq. (1).

Meanwhile, the screened Coulomb potential of the silicon atom is modeled by a sum of Yukawa potentials fitted to the results of the Dirac-Hartree-Fock-Slater (DHFS) self-consistent calculations[15,91]. The full potential generated by all atoms is

$$\phi(\mathbf{r}) = \sum_{i,j} \frac{-Z_j e}{4\pi\varepsilon_0 |\mathbf{r} - \mathbf{r}_i - \mathbf{R}_j|} \sum_k C_{jk} e^{-\mu_{jk}\frac{|\mathbf{r}-\mathbf{r}_i-\mathbf{R}_j|}{a_0}}$$

(5)

where indices $i, j, k$ run over the unit cells, the atoms inside one unit cell, and the fitting parameters of the Yukawa potentials, respectively, $\mathbf{r}_i$ is the coordinate of the $i$-th unit cell, $\mathbf{R}_j$ is the coordinate of the atom with index $j$ inside one unit cell, $Z_j$ is the atomic number, $a_0$ is the Bohr radius, $C_{jk}$ and $\mu_{jk}$ are the fitting parameters of different atoms[91]. For silicon, $C_1 = 0.5160$, $C_2 = 0.4840$, $\mu_1 = 5.8492$ and $\mu_2 = 1.1732$. During the CB process, the electron experiences transverse and inelastic recoils induced by the Coulomb potential and the energy loss from emitting photons, respectively.

**Scattering element of CXR**

This section describes the scattering matrix elements of CXR, used for calculating Eq. (3) in the quantum framework, combining contributions from PXR and CB.

PXR and CB are first and second-order processes in quantum electrodynamics[81], with the corresponding scattering element $M^{PXR}_{\mathbf{q}_i s_i \to \mathbf{q}_f s_f, \mathbf{k}\sigma} = \frac{\delta_{s_i,s_f}}{i\hbar} \int_{-\infty}^{\infty} dt \langle 1_{\mathbf{q}_f s_f} 1_{\mathbf{k}\sigma} | H_{e,ph}(t) | 1_{\mathbf{q}_i s_i} \rangle$ and $M^{CB}_{\mathbf{q}_i s_i \to \mathbf{q}_f s_f, \mathbf{k}\sigma} = \frac{\delta_{s_i,s_f}}{(i\hbar)^2} \int_{-\infty}^{\infty} dt_1 \int_{-\infty}^{t_1} dt_2 \langle 1_{\mathbf{q}_f s_f} 1_{\mathbf{k}\sigma} | H_{e,c}(t_1) H_{e,ph}(t_2) + H_{e,ph}(t_1) H_{e,c}(t_2) | 1_{\mathbf{q}_i s_i} \rangle$, respectively. This calculation neglects the electron spin flip and employs the approximation $\mathbf{u}_{\mathbf{q}_f,s} \approx \mathbf{u}_{\mathbf{q}_i,s}$, which is a very good approximation for a wide range of photon energies including the optical and X-ray spectral ranges. Due to the periodic nature of the interaction Hamiltonians $H_{e,ph}$ and $H_{e,c}$, the scattering elements can be written as $M^{PXR/CB}_{\mathbf{q}_i s_i \to \mathbf{q}_f s_f, \mathbf{k}\sigma} = \sum_{\mathbf{g}} M'^{PXR/CB}_{\mathbf{q}_i s_i \mathbf{g} \to \mathbf{q}_f s_f, \mathbf{k}\sigma} \delta_{\mathbf{q}_i+\mathbf{g},\mathbf{q}_f+\mathbf{k}}$, where momentum conservation is enforced by the Kronecker delta. Details of the derivation can be found in Supplementary Section 3.

**Radiation spectra correction by transverse recoil pathways**

This section derives the unique spectral features of CXR: the split and shift of the radiation spectra, which arise from transverse recoils in the context of energy-momentum conservation. To simplify the analysis without losing of generality, we confine the radiation processes to the $x - z$ plane, and obtain the energy-momentum conservation equations

$$q_{iz} + g_z = q_{fz} + \frac{\omega}{c}\cos\theta$$
$$g_x = q_{fx} + \frac{\omega}{c}\sin\theta \qquad (6)$$
$$E_i(q_{iz}) = E_f(q_{fx}, q_{fz}) + \hbar\omega,$$

where $q_{iz}$ is the incident electron wavevector, $q_{fx}$ and $q_{fz}$ are the $x$ and $z$ components of the final electron wavevector, $g_x$ and $g_z$ are the $x$ and $z$ components of one reciprocal lattice vector, $\theta$ is the radiation angle relative to the electron velocity, $E_i$ and $E_f$ are the energies of the incident and final electrons, $\omega$ is the radiation angular frequency. The radiation angular frequency, derived from Eq. (6) under the approximation of small recoil, is implicitly defined by the following equations

$$\begin{aligned} \omega &= \frac{\hbar q_{iz}}{\gamma m}(q_{iz} - q_{fz}) - \frac{\hbar}{2\gamma m}(q_{iz} - q_{fz})^2 - \frac{\hbar}{2\gamma m}q_{fx}^2 \\ q_{fx} &= g_x - \frac{\omega}{c}\sin\theta \\ q_{iz} - q_{fz} &= -g_z + \frac{\omega}{c}\cos\theta, \end{aligned} \quad (7)$$

where $\gamma$ is the Lorentz factor.

We adopt the linearized electron dispersion approximation by retaining only the first order of electron wavevector changes. Then, Eq. (7) reduces to the classical dispersion equation,

$$\omega_c = v\left(-g_z + \frac{\omega_c}{c}\cos\theta_c\right), \quad (8)$$

where $v = \frac{\hbar q_{iz}}{\gamma m}$ is the velocity of the incident electron, and the subscript c is used to denote the classical value. However, by retaining up to the second order of electron wavevector changes, we obtain the quantum-corrected radiation spectra with the frequency $\omega = \omega_c + \delta\omega$ and radiation direction $\theta = \theta_c + \delta\theta$, as plotted in Fig. 2. The quantum corrected values $\delta\omega$ and $\delta\theta$ are connected by

$$\begin{aligned} \delta\omega(1 - \beta\cos\theta_c) &+ \left(\delta\theta\sin\theta_c + \frac{\delta\theta^2}{2}\cos\theta_c\right)\beta\omega_c \\ &\approx -\frac{\hbar}{2\gamma m}\left[(q_{iz} - q_{fz})^2 + q_{fx}^2\right] \\ &\approx -\frac{\hbar}{2\gamma m}\left[\frac{\omega_c^2}{v^2} + \left(g_x - \frac{\omega_c}{c}\sin\theta_c\right)^2\right], \end{aligned} \quad (9)$$

We retain the second order of $\delta\theta$ to include the case where $\sin\theta_c = 0$.

Combining Eqs. (8) and (9), we note that the photon energy, which causes the electron to be inelastically recoiled, is primarily tied to the longitudinal reciprocal lattice vector component $g_z$, and is only weakly affected by the transverse reciprocal lattice vector components $g_x$. However, the electron transverse recoils, denoted by $q_{fx}$ in Eq. (7), are mainly affected by $g_x$.

Multiple $g_x$ components generate different pathways of electron transverse recoil, and meanwhile, shift and split the radiation spectra as outlined by Eq. (9). By keeping $\delta\omega$ constant under two different transverse reciprocal lattice vector components, $g_x = 0$ and $g'_x$, Eq. (9) corresponds to the illustration shown in Fig. 1c.

## Reference

1. García de Abajo, F. J. Optical excitations in electron microscopy. *Rev. Mod. Phys.* **82**, 209–275 (2010).

2. Pellegrini, C., Marinelli, A. & Reiche, S. The physics of x-ray free-electron lasers. *Rev. Mod. Phys.* **88**, 015006 (2016).

3. Gover, A. *et al.* Superradiant and stimulated-superradiant emission of bunched electron beams. *Rev. Mod. Phys.* **91**, 035003 (2019).

4. Barwick, B., Flannigan, D. J. & Zewail, A. H. Photon-induced near-field electron microscopy. *Nature* **462**, 902–906 (2009).

5. García de Abajo, F. J., Asenjo-Garcia, A. & Kociak, M. Multiphoton Absorption and Emission by Interaction of Swift Electrons with Evanescent Light Fields. *Nano Lett.* **10**, 1859–1863 (2010).

6. Park, S. T., Lin, M. & Zewail, A. H. Photon-induced near-field electron microscopy (PINEM): theoretical and experimental. *New J. Phys.* **12**, 123028 (2010).

7. Huang, S. *et al.* Quantum recoil in free-electron interactions with atomic lattices. *Nat. Photon.* **17**, 224–230 (2023).

8. Davisson, C. & Germer, L. H. The Scattering of Electrons from Single Crystals of Nickel. *Nature* **119**, 558–560 (1927).

9. Kapitza, P. L. & Dirac, P. A. M. The reflection of electrons from standing light waves. *Math. Proc. Camb. Philos. Soc.* **29**, 297–300 (1933).

10. Freimund, D. L., Aflatooni, K. & Batelaan, H. Observation of the Kapitza–Dirac effect. *Nature* **413**, 142–143 (2001).

11. Batelaan, H. Colloquium: Illuminating the Kapitza-Dirac effect with electron matter optics. *Rev. Mod. Phys.* **79**, 929–941 (2007).

12. García de Abajo, F. J., Barwick, B. & Carbone, F. Electron diffraction by plasmon waves. *Phys. Rev. B* **94**, 041404 (2016).

13. Tsesses, S. *et al.* Tunable photon-induced spatial modulation of free electrons. *Nat. Mater.* **22**, 345–352 (2023).

14. Broglie, L. De. Recherches sur la théorie des Quanta (Researches on the quantum theory). *Ann. Phys.* **10**, 22–128 (1925).

15. Wong, L. J. *et al.* Control of quantum electrodynamical processes by shaping electron wavepackets. *Nat. Commun.* **12**, 1700 (2021).

16. Karnieli, A., Rivera, N., Arie, A. & Kaminer, I. The coherence of light is fundamentally tied to the quantum coherence of the emitting particle. *Sci. Adv.* **7**, eabf8096 (2021).

17. Ben Hayun, A. *et al.* Shaping quantum photonic states using free electrons. *Sci. Adv.* **7**, eabe4270 (2021).

18. Mechel, C. *et al.* Quantum correlations in electron microscopy. *Optica* **8**, 70–78 (2021).

19. Lim, J., Kumar, S., Ang, Y. S., Ang, L. K. & Wong, L. J. Quantum Interference between Fundamentally Different Processes Is Enabled by Shaped Input Wavefunctions. *Adv. Sci.* **10**, 2205750 (2023).

20. Konečná, A., Iyikanat, F. & García de Abajo, F. J. Entangling free electrons and optical excitations. *Sci. Adv.* **8**, eabo7853 (2022).

21. Kfir, O. Entanglements of Electrons and Cavity Photons in the Strong-Coupling Regime. *Phys. Rev. Lett.* **123**, 103602 (2019).

22. Di Giulio, V., Kociak, M. & García de Abajo, F. J. Probing quantum optical excitations with fast electrons. *Optica* **6**, 1524–1534 (2019).

23. Gover, A. & Yariv, A. Free-Electron--Bound-Electron Resonant Interaction. *Phys. Rev. Lett.* **124**, 64801 (2020).

24. Ruimy, R., Gorlach, A., Mechel, C., Rivera, N. & Kaminer, I. Toward Atomic-Resolution Quantum Measurements with Coherently Shaped Free Electrons. *Phys. Rev. Lett.* **126**, 233403 (2021).

25. Zhao, Z., Sun, X.-Q. & Fan, S. Quantum Entanglement and Modulation Enhancement of Free-Electron--Bound-Electron Interaction. *Phys. Rev. Lett.* **126**, 233402 (2021).

26. Dahan, R. *et al.* Imprinting the quantum statistics of photons on free electrons. *Science* **373**, eabj7128 (2022).

27. Dahan, R. *et al.* Creation of Optical Cat and GKP States Using Shaped Free Electrons. *Phys. Rev. X* **13**, 31001 (2023).

28. Ginzburg, V. L. Quantum theory of radiation of electron uniformly moving in medium. *Zh. Eksp. Teor. Fiz* **10**, 589–600 (1940).

29. Kaminer, I. *et al.* Quantum Čerenkov Radiation: Spectral Cutoffs and the Role of Spin and Orbital Angular Momentum. *Phys. Rev. X* **6**, 011006 (2016).

30. Tsesses, S., Bartal, G. & Kaminer, I. Light generation via quantum interaction of electrons with periodic nanostructures. *Phys. Rev. A* **95**, 013832 (2017).

31. Mikhail L Ter-Mikhaelyan. Electromagnetic radiative processes in periodic media at high energies. *Physics-Uspekhi* **44**, 571 (2001).

32. Baryshevsky, V. G. *et al.* Coherent bremsstrahlung and parametric X-ray radiation from nonrelativistic electrons in a crystal. *Tech. Phys. Lett.* **32**, 392–395 (2006).

33. Shentcis, M. *et al.* Tunable free-electron X-ray radiation from van der Waals materials. *Nat. Photon.* **14**, 686–692 (2020).

34. Blazhevich, S. V *et al.* First observation of interference between parametric X-ray and coherent bremsstrahlung. *Phys. Lett. A* **195**, 210–212 (1994).

35. Kleiner, V. L., Nasonov, N. N. & Safronov, A. G. Interference between Parametric and Coherent Bremsstrahlung Radiation Mechanisms of a Fast Charged Particle in a Crystal. *Phys. status solidi* **181**, 223–231 (1994).

36. Morokhovskyi, V., Freudenberger, J., Genz, H., Morokhovskii, V. & Richter, A. Theoretical description and experimental detection of the interference between parametric X radiation and coherent bremsstrahlung. *Phys. Rev. B* **61**, 3347–3352 (2000).

37. Feranchuk, I. D., Ulyanenkov, A., Harada, J. & Spence, J. C. H. Parametric x-ray radiation and coherent bremsstrahlung from nonrelativistic electrons in crystals. *Phys. Rev. E* **62**, 4225–4234 (2000).

38. Garybyan, G. M. & Yang, C. Quantum microscopic theory of radiation by a charged particle moving uniformly in a crystal. *Sov. Phys. JETP* **34**, 495 (1972).

39. Baryshevsky, V. G. & Feranchuk, I. D. Transition radiation of γ rays in a crystal. *Sov. Phys. JETP* **34**, 502–504 (1972).

40. Feranchuk, I. D. & Ivashin, A. V. Theoretical investigation of the parametric X-ray features. *J. Phys.* **46**, 1981–1986 (1985).

41. Williams, E. J. Correlation of certain collision problems with radiation theory. *Kong. Dan. Vid. Sel. Mat. Fys. Med.* **13N4**, 1–50 (1935).

42. Überall, H. High-Energy Interference Effect of Bremsstrahlung and Pair Production in Crystals. *Phys. Rev.* **103**, 1055–1067 (1956).

43. Wong, L. J. & Kaminer, I. Prospects in x-ray science emerging from quantum optics and nanomaterials. *Appl. Phys. Lett.* **119**, 130502 (2021).

44. Sandström, S. E. & Überall, H. Channeling radiation and coherent bremsstrahlung for simple lattices: A three-dimensional approach. *Phys. Rev. B* **43**, 12701–12706 (1991).

45. Lohmann, D. *et al.* Linearly polarized photons at MAMI (Mainz). *Nucl. Instrum. Meth. A* **343**, 494–507 (1994).

46. Morokhovskii, V. V. *et al.* Polarization of parametric X radiation. *Phys. Rev. Lett.* **79**, 4389–4392 (1997).

47. Shchagin, A. V. Linear polarization of parametric X-rays. *Phys. Lett. A* **247**, 27–36 (1998).

48. Schmidt, K. H. *et al.* Measurement of the linear polarization of Parametric X-radiation. *Nucl. Instrum. Meth. B* **145**, 8–13 (1998).

49. Talebi, N. Electron-light interactions beyond the adiabatic approximation: Recoil engineering and spectral interferometry. *Adv. Phys. X* **3**, 1499438 (2018).

50. Feist, A., Yalunin, S. V, Schäfer, S. & Ropers, C. High-purity free-electron momentum states prepared by three-dimensional optical phase modulation. *Phys. Rev. Res.* **2**, 43227 (2020).

51. Dahan, R. *et al.* Resonant phase-matching between a light wave and a free-electron wavefunction. *Nat. Phys.* **16**, 1123–1131 (2020).

52. Tsarev, M., Thurner, J. W. & Baum, P. Nonlinear-optical quantum control of free-electron matter waves. *Nat. Phys.* **19**, 1350–1354 (2023).

53. Liu, F. *et al.* Integrated Cherenkov radiation emitter eliminating the electron velocity threshold. *Nat. Photon.* **11**, 289–292 (2017).

54. Roques-Carmes, C. *et al.* Towards integrated tunable all-silicon free-electron light sources. *Nat. Commun.* **10**, 3176 (2019).

55. Ye, Y. *et al.* Deep-ultraviolet Smith–Purcell radiation. *Optica* **6**, 592–597 (2019).

56. Zhang, D. *et al.* Coherent surface plasmon polariton amplification via free-electron pumping. *Nature* **611**, 55–60 (2022).

57. Smith, S. J. & Purcell, E. M. Visible light from localized surface charges moving across a grating. *Phys. Rev.* **92**, 1069 (1953).

58. Ritchie, R. H. Plasma losses by fast electrons in thin films. *Phys. Rev.* **106**, 874 (1957).

59. Ginzburg, V. L. Radiation by uniformly moving sources (Vavilov-Cherenkov effect, transition radiation, and other phenomena). *Phys. Usp.* **39**, 973–982 (1996).

60. Adamo, G. *et al.* Light well: A tunable free-electron light source on a chip. *Phys. Rev. Lett.* **103**, 113901 (2009).

61. Liu, S. *et al.* Surface Polariton Cherenkov Light Radiation Source. *Phys. Rev. Lett.* **109**, 153902 (2012).

62. Peralta, E. A. *et al.* Demonstration of electron acceleration in a laser-driven dielectric microstructure. *Nature* **503**, 91–94 (2013).

63. Wong, L. J., Kaminer, I., Ilic, O., Joannopoulos, J. D. & Soljačić, M. Towards graphene plasmon-based free-electron infrared to X-ray sources. *Nat. Photon.* **10**, 46–52 (2016).

64. Remez, R. *et al.* Observing the Quantum Wave Nature of Free Electrons through Spontaneous Emission. *Phys. Rev. Lett.* **123**, 060401 (2019).

65. García de Abajo, F. J. & Di Giulio, V. Optical Excitations with Electron Beams: Challenges and Opportunities. *ACS Photonics* **8**, 945–974 (2021).

66. Lin, K. *et al.* Ultrafast Kapitza-Dirac effect. *Science* **383**, 1467–1470 (2024).

67. Feist, A. *et al.* Quantum coherent optical phase modulation in an ultrafast transmission electron microscope. *Nature* **521**, 200–203 (2015).

68. Madey, J. M. J. Stimulated Emission of Bremsstrahlung in a Periodic Magnetic Field. *J. Appl. Phys.* **42**, 1906–1913 (1971).

69. Becker, W. & McIver, J. K. Quantum descriptions of free-electron lasers. *J. Phys. Colloq.* **44**, C1-289-C1-311 (1983).

70. Becker, W. & Zubairy, M. S. Photon statistics of a free-electron laser. *Phys. Rev. A* **25**, 2200–2207 (1982).

71. Bonifacio, R., Piovella, N. & Robb, G. R. M. The quantum free electron laser: A new source of coherent, short-wavelength radiation. *Fortschr. Phys.* **57**, 1041–1051 (2009).

72. Kling, P. *et al.* What defines the quantum regime of the free-electron laser? *New J. Phys.* **17**, 123019 (2015).

73. Bambini, A., Renieri, A. & Stenholm, S. Classical theory of the free-electron laser in a moving frame. *Phys. Rev. A* **19**, 2013–2025 (1979).

74. Reese, G. M., Spence, J. C. H. & Yamamoto, N. Coherent bremsstrahlung from kilovolt electrons in zone axis orientations. *Philos. Mag. A* **49**, 697–716 (1984).

75. Andersen, J. U. & Lægsgaard, E. Coherent bremsstrahlung and sidebands for channeled electrons. *Nucl. Instrum. Meth. B* **33**, 11–17 (1988).

76. Kurizki, G. Bloch waves and band structure for diffracted and channeled particles in crystals.

*Phys. Rev. B* **33**, 49–63 (1986).

77. Überall, H. & Sáenz, A. W. Channeling radiation and coherent bremsstrahlung. *Phys. Lett. A* **90**, 370–374 (1982).

78. Lin, X. *et al.* Controlling Cherenkov angles with resonance transition radiation. *Nat. Phys.* **14**, 816–821 (2018).

79. Balanov, A., Gorlach, A. & Kaminer, I. Temporal and spatial design of x-ray pulses based on free-electron-crystal interaction. *APL Photonics* **6**, 070803 (2021).

80. Huang, S. *et al.* Enhanced Versatility of Table-Top X-Rays from Van der Waals Structures. *Adv. Sci.* **9**, 2105401 (2022).

81. Baryshevsky, V. G., Feranchuk, I. D. & Ulyanenkov, A. P. *Parametric X-Ray Radiation in Crystals: Theory, Experiment and Applications*. (Springer, 2005).

82. Rivera, N. & Kaminer, I. Light–matter interactions with photonic quasiparticles. *Nat. Rev. Phys.* **2**, 538–561 (2020).

83. Wong, L. W. W. *et al.* Free-electron crystals for enhanced X-ray radiation. *Light Sci. Appl.* **13**, 29 (2024).

84. Guzzinati, G. *et al.* Probing the symmetry of the potential of localized surface plasmon resonances with phase-shaped electron beams. *Nat. Commun.* **8**, 14999 (2017).

85. Verbeeck, J. *et al.* Demonstration of a 2 × 2 programmable phase plate for electrons. *Ultramicroscopy* **190**, 58–65 (2018).

86. Shi, X. *et al.* Free-electron-driven X-ray caustics from strained van der Waals materials. *Optica* **10**, 292–301 (2023).

87. Shi, X. *et al.* Free-electron interactions with van der Waals heterostructures: a source of focused X-ray radiation. *Light Sci. Appl.* **12**, 148 (2023).

88. Roques-Carmes, C. *et al.* Free-electron–light interactions in nanophotonics. *Appl. Phys. Rev.* **10**, 011303 (2023).

89. Donnelly, C. *et al.* Three-dimensional magnetization structures revealed with X-ray vector nanotomography. *Nature* **547**, 328–331 (2017).

90. Authier, A. *Dynamical Theory of X-ray Diffraction*. (Oxford University Press, 2001).

91. Salvat, F., Martnez, J. D., Mayol, R. & Parellada, J. Analytical Dirac-Hartree-Fock-Slater screening function for atoms (Z=192). *Phys. Rev. A* **36**, 467 (1987).